\documentstyle[12pt,epsfig]{article}
\begin{document}

\begin{center}
{\Large {\bf The event generator DECAY4 for simulation of double beta
processes and decay of radioactive nuclei}}

\vskip 0.4cm {\bf O.A.~Ponkratenko, V.I.~Tretyak and Yu.G.~Zdesenko }

{\it Institute for Nuclear Research, Prospect Nauki 47, MSP 03680 Kiev,
Ukraine}
\end{center}

\begin{abstract}
\noindent
The computer code DECAY4 is developed to generate initial energy, time
and angular distributions of particles emitted in 
radioactive decays of nuclides and nuclear (atomic) deexcitations. Data
for description of nuclear and atomic decay schemes are taken from the
ENSDF and EADL database libraries. The examples of use of the DECAY4 code in
several underground experiments are described.
\end{abstract}

\section{Introduction and overall description of the DECAY4}

\indent\indent
Despite the fact that effect to background ratio is the central problem of
all experimental physics, there is a certain class of experiments for which
this problem is so crucial that even the possibility of their performance
itself depends strongly on the reached background level of the used
detectors. These are so called underground experiments devoted to
investigation of the very rare or forbidden decays and processes like, for
instance, double beta decay, proton decay, dark matter particles search,
solar neutrino study and so on. The ultimate sensitivity of such experiments
is determined mainly (except the available source strengths) by the detector
background. The first origin of background is due to cosmic rays and can be
eliminated by the proper underground site for the set up. The second (and
the most crucial for sensitivity) source of the background is the decays of
the nuclides from the radioactive impurities in the detector itself, in the
materials used for detector mounting and shielding, and in the surroundings.
Therefore it is obvious that simulation of the background and, in
particular, simulation of the nuclides decays is the overwhelmingly
important part of such kind of research which can allow: i) to understand
and determine the origins of the background and hence to find certain
methods to eliminate or suppress the background contributions; ii) to build
up the background model and response functions of the detector for the
effect being sought (together with the detector energy and efficiency
calibrations, resolution, source activities, etc.), thus to extract and
evaluate searched effect (or to exclude it) more precisely.

There are several general programs, which are commonly used for simulation
of the particles interactions in the experimental set up as, for example,
GEANT package \cite{GEANT} or EGS4 code \cite{EGS4}. In any such a program
user should describe initial kinematics of events by using the so called
event generator. The last is an important part of the simulation program
providing the information, which particles and how many of them are emitted,
what are their energies, directions of movement and times of emission.
Existing computer codes RADLST \cite{RADLST} and IMRDEC \cite{IMRDEC} only
determine the radiation spectra due to decay of nuclides, and therefore
could not be used for the further particles tracking.

In attempt to cover this lack, the code, named DECAY4, was developed for
generation of events in the low energy nuclear and particle physics (double
beta decay and decay of radioactive nuclides). This code was elaborated
during the last decade, mainly for 2$\beta $ decay research \cite{Zde89}.
The plan of the paper is as following: first, the overall features of the
DECAY4 generator and used databases are considered, then detail descriptions
of the parts associated with the double beta decay of atomic nuclei and
decays of the natural and artificial radioactive nuclides are presented. In
the last section several examples illustrating the use of the DECAY4
generator in the real underground experiments are shown.

The developed program DECAY4 gives the possibility to generate events of the
2$\beta $ decay of atomic nuclei and of the radioactive decays ($\alpha $, $%
\beta ^{\pm },$ $p,$ $n$ decay, electron capture) of all known unstable
isotopes. It is divided into two main sections:

a) INIT -- search and reading of all parameters of the nucleus and its decay
needed for decay simulation from the ENSDF \cite{ENSDF} (or NuDat \cite
{NuDat}), EADL \cite{EADL} and other libraries, in order to build up the
nuclear and atomic decay schemes;

b) GENDEC -- a Monte Carlo events generator itself.

The ENSDF database library includes the following information on about 2500
isotopes used for generation of the radioactive decays: a) decay modes,
their probabilities and energy releases, isotopes half-lives; b) radiation
type, particles energies and intensities; c) parameters of nuclear levels
(half-life, spin, parity and excitation energy); d) parameters of nuclear
transitions (branching ratios, multipolarities, coefficients of internal
conversion and mixing ratios).

The DECAY4 code uses also tables with the data on atomic properties of
isotopes (electron binding energies, electron capture (EC) subshell ratios,
X rays and Auger electrons intensities) from the EADL database \cite{EADL}
(as well as from \cite{TOI8}) and tables with theoretical Hager-Seltzer
conversion coefficients \cite{HSICC}.

The GENDEC part of the DECAY4 generates the energy, time of emission,
direction and polarization for the following emitted particles: 1) electrons
and positrons from single and double $\beta $ decay; 2) $\alpha $ particles
from $\alpha $ decay, protons and neutrons from $p$ and $n$ decays; 3) $%
\gamma $ quanta from nuclear deexcitation process; 4) conversion electrons;
5) e$^{-}$e$^{+}$ pairs from the internal pair conversion; 6) $\gamma $
quanta from bremsstrahlung in $\beta $ decay and EC; 7) neutrinos
(antineutrinos) from EC or $\beta $ (2$\beta $) decay; 8) X rays and Auger
electrons from the atomic deexcitation process.

\section{Double beta decay processes}

\indent\indent
The DECAY4 describes double beta processes (2$\beta ^{-}$ and 2$\beta ^{+}$
decays, electron capture with emission of positron $\varepsilon \beta ^{+}$
and double electron capture 2$\varepsilon $) for all nuclides. 2$\beta $
transitions to the ground state as well as to excited 0$^{+}$ and 2$^{+}$
levels of the daughter nucleus are allowed. If 2$\beta $ decay occurs to an
excited level of a nucleus, the electromagnetic deexcitation process
follows. The energy release $Q_{\beta \beta }$ for double beta processes is
taken from the table of the atomic masses \cite{Aud95}. For each transition
to the ground or an excited level, various modes (with emission of two
neutrinos or Majoron, neutrinoless decays due to nonzero neutrino mass or
right-handed admixture in the weak interaction, etc.) and mechanisms
(two-nucleon $2n$ and $\Delta $-isobar N$^{*}$) of double beta decay are
possible. Below we give the list of double beta processes which could be
simulated with the DECAY4:

1.{\it {\bf \ $0\nu 2\beta ^{\pm }$ }}decay with neutrino mass{\it {\bf , $%
0^{+}-0^{+}$ }}transition{\it {\bf , $2n$-}}mechanism;

2.{\it {\bf \ $0\nu 2\beta ^{\pm }$ }}decay with right-handed currents,{\it 
{\bf \ $0^{+}-0^{+}$ }}transition{\it {\bf , $2n$-}}mechanism;

3.{\it {\bf \ $0\nu 2\beta ^{\pm }$ }}decay with right-handed currents,{\it 
{\bf \ $0^{+}-0^{+}$ }}and{\it {\bf \ $0^{+}-2^{+}$ }}transitions{\it {\bf , 
}}N{\it {\bf $^{*}$-}}mechanism;

4.{\it {\bf \ $2\nu 2\beta ^{\pm }$ }}decay,{\it {\bf \ $0^{+}-0^{+}$ }}%
transition,{\it {\bf \ $2n$-}}mechanism;

5.{\it {\bf \ $0\nu 2\beta ^{\pm }$ }}decay with emission of Majoron,{\it 
{\bf \ $0^{+}-0^{+}$ }}transition{\it {\bf , $2n$-}}mechanism;

6.{\it {\bf \ $0\nu 2\beta ^{\pm }$ }}decay with double Majoron emission{\it 
{\bf , $0^{+}-0^{+}$ }}transition,{\it {\bf \ $2n$-}}mechanism; decay with
charged{\it {\bf \ $L=-2$ }}Majoron or massive vector Majoron;

7.{\it {\bf \ $0\nu 2\beta ^{\pm }$ }}decay with right-handed currents,{\it 
{\bf \ $0^{+}-2^{+}$}} transition, {\it {\bf $2n$-}}mechanism;

8.{\it {\bf \ $2\nu 2\beta ^{\pm }$ }}decay,{\it {\bf \ $0^{+}-2^{+}$ }}%
transition,{\it {\bf \ $2n$- }}and N{\it {\bf $^{*}$-}}mechanisms;

9.{\it {\bf \ $0\nu $}}$\varepsilon ${\it {\bf $\beta ^{+}$ }}decay;

10.{\it {\bf \ $2\nu $}}$\varepsilon ${\it {\bf $\beta ^{+}$ }}decay{\it 
{\bf , $0^{+}-0^{+}$ }}and{\it {\bf \ $0^{+}-2^{+}$ }}transitions;

11.{\it {\bf \ $0\nu 2$}}$\varepsilon ${\it {\bf \ }}decay;

12.{\it {\bf \ $2\nu 2$}}$\varepsilon ${\it {\bf \ }}decay,{\it {\bf \ $%
0^{+}-0^{+}$ }}and{\it {\bf \ $0^{+}-2^{+}$ }}transitions.

The theoretical formulae for the energy and angular distribution $\rho
(E_1,E_2,cos\theta )$ of emitted electrons or positrons are based on works 
\cite{Doi81, Moh88, Bur93, Car93, Tre95}. For example, for the first process$%
:$%
\begin{equation}
\rho (E_1,E_2,cos\theta )=p_1(E_1+1)F(E_1,Z)~p_2(E_2+1)F(E_2,Z)\delta
(E_0-E_1-E_2)(1-\beta _1\beta _2cos\theta ), 
\end{equation}
where $E_i$ is the kinetic energy of the $i$-th e$^{-}$ or e$^{+}$ (in units
of the electron mass $m_ec^2$), $p_i$ is the momentum (in units of $m_ec$), $%
F(E_i,Z)$ is the Fermi function, $Z$ is the atomic number of the daughter
nucleus ($Z>0$ for 2$\beta ^{-}$ and $Z<0$ for 2$\beta ^{+}$ decay), $\theta 
$ is the angle between the particles directions, $E_0$ is the energy
available for the particles ($E_0=Q_{\beta \beta }-E_j^{ex}$ for 2$\beta
^{-} $ decay and $E_0=Q_{\beta \beta }-4-E_j^{ex}$ for 2$\beta ^{+}$ decay, $%
E_j^{ex}$ is the energy of the populated level of the daughter nucleus), $%
\beta _i=p_i/(E_i+1).$

\section{Radioactive decays of nuclides}

\indent\indent
The DECAY4 describes six decay modes: $\beta ^{-}$, $\alpha ,$ $p$ and $n$
decays, electron capture and $\beta ^{+}$ decay (EC) and isomeric transition
(IT). The modes $d$ ($d=\beta ^{-}$, $\alpha $, $p$, $n$, EC, IT), their
probabilities $p^d,$ available decay energies $Q^d$ and isotopes half-life $%
T_{1/2}$ are taken from the ENSDF \cite{ENSDF} or NuDat \cite{NuDat}
databases. The decay mode $d$ is sampled according to the probabilities 
$p^d$.

${\bf \beta }^{-}$ {\bf decay.} The endpoint energies in $\beta ^{-}$ decay $%
E_i^0$ are related with the energy release $Q^{\beta ^{-}}$ and the level
energies $E_i^{ex}$ of the daughter nucleus by the equation 
\begin{equation}
Q^{\beta ^{-}}=E_i^0+E_i^{ex}.  
\end{equation}

Kinetic energy of the beta particle $E$ is sampled in accordance with the
distribution 
\begin{equation}
\rho \left( E\right) =p\cdot \left( E+1\right) \cdot \left( E_i^0-E\right)
^2\cdot F(E,Z)\cdot S_k\left( E\right) , 
\end{equation}
where $S_k\left( E\right) $ is the factor of forbiddenness. Probability of
the internal bremsstrahlung in beta decay and energy-angular distribution of
bremsstrahlung $\gamma $ quanta are calculated as in \cite{ABGS1}.

${\bf \alpha }$, ${\bf p}${\bf , }${\bf n}$ {\bf decays.} The energies of
particles $E_i^k$ ($k=\alpha ,$ $p,$ $n$) are related with the level
energies $E_i^{ex}$ of the daughter nucleus by the equation 
\begin{equation}
E_i^k=A_d/A_p\cdot (Q^k-E_i^{ex}), 
\end{equation}
where $A_p,$ $A_d$ is the mass numbers of the parent and daughter nuclei,
respectively.

{\bf EC\ (electron capture\ and\ }${\bf \beta }^{{\bf +}}$ {\bf decay).} The
ENSDF database includes the information on the probabilities $p_i^{EC}$ (for
EC) and $p_i^{\beta ^{+}}$ (for $\beta ^{+}$ decay) for $i$-th level of the
daughter nucleus. If the level is populated in $\beta ^{+}$ decay, energy of
positron is sampled according to (3), where 
\begin{equation}
E_i^0=Q^{EC}-2-E_i^{ex}.  
\end{equation}

If the $i$-th nuclear level was populated in the electron capture, the
number of the atomic subshell $x$ ($x=K,$ $L_1,$ $L_{2,}$ $L_{3,}$ $M_1,$ $%
M_2,$ $...,$ $M_5,$ $N_1,$ $N_2,$ $...,$ $N_7,$ $O_1,$ $O_2,$ $...,$ $O_7$),
where primary electron vacancy is created, is sampled according to
probabilities $P_x^{EC}$ \cite{TOI8}: 
\begin{equation}
P_x^{EC}(Z,q_x)=const\cdot n_xp_x^{2(k_x-1)}q_x^{2(L-k_x+1)}\beta
_x^2B_x/[(2k_x-1)!(2L-2k_x+1)!],  
\end{equation}
where $L$ is the electron capture transition angular momentum, $n_x$ is the
relative occupation number for partially filled subshells $x$ ($%
n_x=N_x/N_x^{\max },$ $N_x$ is the number of electrons in the subshell $x$, $%
N_x^{\max }$ is the maximal number of electrons in the subshell), $%
q_x=Q^{EC}-E_i^{ex}-E_x$ is the neutrino energy, $E_x$ is the electron
binding energy in the parent atom, and $k_x$ is $x$ subshell angular
momentum. The squared amplitudes $\beta _x^2B_xp_x^{2(k_x-1)}$ of the
bound-state electron radial wave functions and the electron binding energy $%
E_x$ are taken from \cite{Bam77, TOI8}. The internal bremsstrahlung
probability and spectra of bremsstrahlung $\gamma $ quanta in allowed
electron capture transition from $x$ atomic subshell is calculated in
accordance with \cite{Bam77}.

{\bf Nuclear deexcitation process.} The nuclear deexcitation process occurs
if a daughter nucleus is in the excited $i$-th level with energy $E_i^{ex}$.
The electromagnetic transition from $i$-th to $j$-th level is sampled
according to probabilities 
\begin{equation}
p_{ij}=J_{ij}/\sum\nolimits_jJ_{ij},  
\end{equation}
where $J_{ij}$ is the branching ratio of electromagnetic transition from $i$%
-th level to $j$-th taken from the NuDat or ENSDF databases.

There are three possible modes of electromagnetic transition with emission
of: (1) $\gamma $ quantum with energy $E^\gamma =E_i^{ex}-E_j^{ex};$ (2)
conversion electron with energy $E_x^{ce}=E_i^{ex}-E_j^{ex}-E_x$ (the
condition $E_x^{ce}>0$ should be fulfilled), here $E_x$ is the electron
binding energy on the $x$ subshell$;$ (3) conversion electron-positron pair
with total energy $E^{cp}=E_i^{ex}-E_j^{ex}-2$ (if $E^{cp}>0$). To sample
the mode, the respective probabilities $p_{ij}^\gamma $, $p_{ij}^{ce}$, $%
p_{ij}^{cp}$ are used, where 
\begin{equation}
p_{ij}^\gamma =1/(1+\alpha _{ij}^{ce}+\alpha _{ij}^{cp}), ~ 
p_{ij}^{ce}=\alpha _{ij}^{ce}p_{ij}^\gamma, ~ 
p_{ij}^{cp}=\alpha_{ij}^{cp}p_{ij}^\gamma ~
\end{equation}
for all transitions beside E0,
\begin{equation}
p_{ij}^\gamma =0, ~
p_{ij}^{ce}=1/(1+I_{ij}^{cp}/I_{ij}^{ce}), ~
p_{ij}^{cp}=I_{ij}^{cp}/I_{ij}^{ce}\cdot p_{ij}^{ce} ~
\end{equation}
for E0 transition,  
where $\alpha _{ij}^{ce}$, $\alpha _{ij}^{cp}$ are the coefficients of the
internal electron and pair conversion, respectively, and $I_{ij}^{ce}$, $%
I_{ij}^{cp}$ are the intensities of internal electron and pair conversion,
respectively.

Total $\alpha _{ij}^{ce}$, partial subshell $\alpha _{ij}^{ce}(s_m)$, shell $%
\alpha _{ij}^{ce}(s)$ coefficients of electron internal conversion is
related by 
\begin{equation}
\alpha _{ij}^{ce}=\sum\nolimits_s\alpha _{ij}^{ce}(s); ~
\alpha_{ij}^{ce}(s)=\sum\nolimits_m\alpha _{ij}^{ce}(s_m),  
\end{equation}
where $s$ is the shell index and $m$ is the subshell index of the $s$ shell $%
(s_m\equiv x)$. Coefficients $\alpha _{ij}^{ce}$ and $\alpha _{ij}^{ce}(s)$
are taken from the NuDat or ENSDF databases. In case if their values are
absent, the coefficients are calculated as 
\begin{equation}
\alpha _{ij}^{ce}(s_m)=[\alpha _{ij}^{ce}(s_m,E^\gamma,
\pi _1\lambda
_1,Z)+\delta _{ij}\alpha _{ij}^{ce}(s_m,E^\gamma ,\pi _2\lambda
_2,Z)]/(1+\delta _{ij}^2),  
\end{equation}
where the values of partial coefficients $\alpha ^{ce}(s_m,E^\gamma ,\pi
\lambda ,Z)$ are taken from \cite{HSICC}. Here $\pi \lambda $ is transition
multipolarity, and $\delta _{ij}$ is the mixing ratio of different
multipolarities in the $i\rightarrow j$ transition. For pure $E0$
transitions, electron conversion coefficients are calculated according to
formulae from \cite{TOI8}. Coefficient of internal pair conversion $\alpha
_{ij}^{cp}$ is given by formulae similar to (10), and partial coefficients $%
\alpha ^{cp}(E^\gamma ,\pi \lambda ,Z)$, and electron-positron energy and
angular distributions are calculated according to formulae from \cite{TOI8,
ROSE}.

{\bf X rays and Auger electrons. }The vacancies in the atomic shells are
created in the electron capture or in the result of internal electron
conversion. Ionized atom is deexcited filling the vacancies by electrons
from higher atomic shells; X rays and Auger electrons are emitted in this
process. For the sampling of the atomic deexcitation process, the values of
the electron binding energies and occupation numbers, radiative and
radiationless partial widths are taken from the EADL library \cite{EADL}.
The type of process (radiation or Auger electron emission) is sampled
accordingly to the values of radiative and radiationless partial widths. For
the vacancy in the $s_i$ subshell, the energy of X ray for radiative
process, in which higher vacancy $q_j$ is created, and the energy of Auger
electron for nonradiative process (vacancies $q_j$ and $t_l$ are created)
are given by 
\begin{equation}
E_{s_i-q_j}^X=E_{s_i}(Z)-E_{q_j}(Z); ~
E_{s_i-q_jt_l}^A=E_{s_i}(Z)-E_{q_j}(Z)-E_{t_l}(Z)-\Delta E_{q_jt_l}. 
\end{equation}
Correction $\Delta E_{q_jt_l}$ is found from equations in \cite{BURH}.
Multivacancy corrections for energies and partial widths are included also.
The X ray and Auger electron energy spreading is accounted according to
Lorentzian distribution function \cite{BURH}.

{\bf Decay chains and time characteristics. }Time interval $T^k$ between
appearance of the unstable state $k$ of the nucleus or atom and its decay
with emission of the particle is sampled in accordance with the relation 
\begin{equation}
T^k=-T_{1/2}^k\ln \eta /\ln 2,  
\end{equation}
where $\eta $ is a random number uniformly distributed in the range $(0,1)$, 
$T_{1/2}^k$ is a half-life of the unstable state. The DECAY4 can also
simulate the full decay chain of the parent nuclide together with its
daughters. As an example, the generated spectra of particles, emitted in
decay of nuclides from full $^{238}$U chain, are shown in fig. 1. Activities
of all daughter nuclei are calculated by the DECAY4.

\nopagebreak
\begin{figure}[ht]
\begin{center}
\mbox{\epsfig{figure=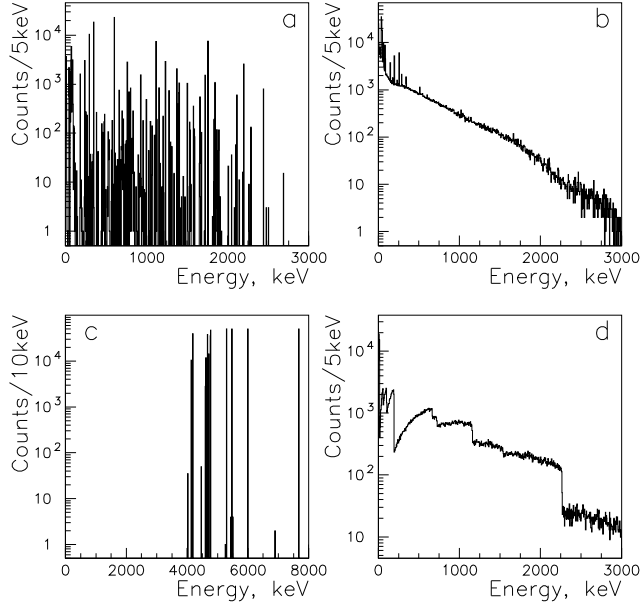,height=8.0cm}}
\caption {Generated initial energy spectra of particles emitted in the
decay of $^{238}$U chain (in equilibrium): (a) $\gamma$ quanta, (b)
electrons, (c) $\alpha$ particles, (d) antineutrinos.}
\end{center}
\end{figure}

{\bf Angular correlations between emitted particles. }The direction (polar
and azimutal angles $\theta _i$, $\varphi _i)$ of the particle $i$ is
sampled as isotropical if one of the next conditions is fulfilled: 1) the
particle is the first in the decay of unoriented parent nuclear state $i$;
2) $I_i<1$ ($I_i$ is nuclear spin before emission of the particle $i);$ 3) $%
T^i>\tau _{\max }$ ($T^i$ is the time given by (13), $\tau _{\max }$ is user
defined parameter (time) to account the influence of external fields which
violate the angular correlation); 4) $\lambda _i$ $=0$ or $\lambda _{i-1}$ $%
=0$ ($\lambda _i$ and $\lambda _{i-1}$ are angular momenta of particles $i$
and $i-1,$ if particle $i-1$ exists)$;$ 5) one of the values $I_{i-1}$, $I_i$%
, $I_{i+1}$, $\lambda _i$, $\lambda _{i-1}$ is unknown and could not be
evaluated. In case, if particle $i+1$ does not satisfy any of the above
conditions but particle $i$ satisfies one of them, the polar angle $\theta
_{i+1}^i$ of particle $i+1$ in the coordinate system of particle $i$ is
sampled in according to correlation function \cite{BIEN} 
\begin{equation}
W(\theta _{i+1}^i)=\sum\nolimits_{k_{i+1}=0}^{k_{i+1}^{\max
}}A_{k_{i+1}}^{-}(\lambda _i\lambda _i^{\prime
}I_{i+1}I_i,x_i)A_{k_{i+1}}^{+}(\lambda _{i+1}\lambda _{i+1}^{\prime
}I_{i+2}I_{i+1},x_{i+1})P_{k_{i+1}}(\cos \theta _{i+1}^i), 
\end{equation}
where $\lambda _i^{\prime }=\lambda _i\pm 1$ is the second possible angular
momentum of the particle (for the mixture of multipolarities), $x_i$ is type
of the particle $i$ ($x_i=\gamma ,$ $\beta ,$ $\alpha ,$ $e),$ $%
A_{k_{i+1}}^{\pm }(\lambda _i\lambda _i^{\prime }I_{i+1}I_i,x_i)$ are
angular momenta functions, and $k$ is even. In this case correlation
function is independent from the azimutal angle $\varphi _{i+1}^i$. If the
next emitted particle $i+2$ exists, which does not satisfy any condition
among 1)--5), its direction is determined by the more complicated
correlation function, which depends on directions of previous particles.
Correlation of linear polarisations of $\gamma $ quanta is also taken into
account \cite{BIEN}.

\section{Conclusions}

\indent\indent
The code DECAY4 was successfully used in several underground experiments for
detectors design and optimization, simulation of the backgrounds (with the
help of the GEANT package) and for evaluation of the results. Some examples
are listed below.

1. Kiev 2$\beta $ decay experiments performed in the Solotvina Underground
Laboratory in a salt mine 430 m underground ($\simeq $1000 m w. e.):

1.1. Cadmium tungstate crystal scintillators (enriched in $^{116}$Cd to
83\%) were used to study $^{116}$Cd \cite{Dan95, Dan99}. The background of
the $^{116}$CdWO$_4$ crystal (15.2 cm$^3$) in the energy region of interest
(Q$_{\beta \beta }$=2805 keV) was equal to $\approx $0.6 counts/y$\cdot $kg$%
\cdot $keV. With the 19175 h statistics the half-life limit for neutrinoless
2$\beta $ decay of $^{116}$Cd was set: T$_{1/2}^{0\nu }\geq 3.2\cdot 10^{22}$
y (90\% C.L.) \cite{Dan99}. Limits on 0$\nu $ modes with emission of one
(M1) or two (M2) Majorons were established also: 
T$_{1/2}^{0\nu M1}\geq 1.2\cdot 10^{21}$ y and 
T$_{1/2}^{0\nu M2}\geq 2.6\cdot 10^{20}$ y (90\% C.L.) \cite{Dan98}. 
Comparing these limits with the theory,
the restrictions on the neutrino mass $\langle m_\nu \rangle \leq 3.9$ eV and
Majoron-neutrino coupling constant $g\leq 2.1\cdot 10^{-4}$ were derived,
which are among the most sensitive results for other nuclei \cite{Tre95}.

1.2. The study of the $2\beta $ decay of $^{160}$Gd was carried out with the
help of the 95 cm$^3$ Gd$_2$SiO$_5$:Ce crystal scintillator. The background
was reduced to $\approx $1.0 cpd/keV$\cdot $kg in the vicinity of Q$_{\beta
\beta }$ energy ($\approx $1.73 MeV). The improved half-life limit was set
for $0\nu 2\beta $ decay of $^{160}$Gd: T$_{1/2}^{0\nu }\geq 1.2\cdot $10$%
^{21}$ y at 68\% C.L. \cite{Gd-96}.

2. DAMA collaboration experiments performed deep underground in the Gran
Sasso National Laboratory:

2.1. Two radiopure CaF$_2$:Eu crystal scintillators (370 g each) have been
used to study $2\beta $ decay of $^{46}$Ca and double electron capture of $%
^{40}$Ca as well as for dark matter search. The highest up-to-date half-life
limits were reached for $0\nu $ and $2\nu $ double electron capture of $%
^{40} $Ca: T$_{1/2}^{0\nu }\geq 4.9\cdot $10$^{21}$ y and T$_{1/2}^{2\nu
}\geq 9.9\cdot $10$^{21}$ y (68\% C.L.) \cite{CaF2-99}.

2.2. The study of the $2\beta ^{+}$ decay of $^{106}$Cd was performed with
the help of two low background NaI(Tl) crystals and enriched (to 68\%) $%
^{106}$Cd samples ($\approx $154 g). New T$_{1/2}$ limits for the $\beta
^{+}\beta ^{+}$, $\beta ^{+}$/EC and EC/EC decay of $^{106}$Cd have been set
in the range of (0.3$-$4)$\cdot $10$^{20}$ y at 90\% C.L. \cite{Cd106-99}%
{\normalsize .}

3. NEMO collaboration experiment performed in the Frejus Underground
Laboratory to study 2$\beta $ decay of $^{100}$Mo \cite{Mo-100}{\normalsize .%
} For background simulation the first version of the DECAY4 code \cite{GENBB}
was used. The clear two neutrino 2$\beta $ signal (1433 events during 6140
h) was observed, leading to a half-life T$_{1/2}=0.95\pm 0.04$(stat)$\pm
0.09 $(syst)$\cdot 10^{19}$ y \cite{Mo-100}{\normalsize .}

4. Heidelberg-Kiev collaboration: The background simulations were performed
for the GENIUS project which aim is to increase the present sensitivity for 2%
$\beta $ decay and dark matter search. Contributions from the cosmogenic
activity produced in the Ge detectors and from their radioactive impurities
as well as from contamination of the liquid nitrogen and other materials
were calculated. External $\gamma $, $\mu $ and neutron backgrounds were
considered also. The results of calculations evidently show feasibility of
the GENIUS experiment \cite{GENIUS}{\normalsize .}

In result, we can conclude:

1. The DECAY4 code is the powerful tool for simulation of the radionuclides
decays in the wide range of decay's modes, emitting particles, etc.

2. The DECAY4 can be connected easily with the codes which simulate the
particles propagation, like GEANT, EGS4, etc.

3. The code includes the most advanced databases: ENSDF, NuDat, EADL and
others.

4. The event generator DECAY4 was used successfully in many underground
experiments (Kiev, Roma-Kiev, DAMA-, NEMO-, GENIUS-collaborations).\\

\indent\indent
{\bf Acknowledgments.} This work was supported in part by the Science and
Technology Center of Ukraine (project \# 411).


\end{document}